\documentstyle[preprint,aps,prl]{revtex}
\begin{document}
\preprint{1}
\draft
\title{ Incipient antiferromagnetism and low-energy excitations
in the half-filled two-dimensional Hubbard model
}
\author{  J.J. Deisz$^a$,  D.W. Hess$^b$, and J.W. Serene$^a$  \\
 \hfill }
\address{
$^a$Department of Physics \\
Georgetown University \\
Washington, D.C. \ \ 20057-0995 \\
\hfill
}
 \address{
$^b$Complex Systems Theory Branch, \\
Naval Research Laboratory, \\
Washington, D.C. \ 20375-5345 \\
}
\vspace{0.25in}
\date{August 4, 1995; slightly revised from March 9, 1995 submission}
\maketitle
\begin{abstract}
We present single-particle and thermodynamic properties
of the half-filled single-band Hubbard model in 2D calculated
in the self-consistent fluctuation exchange approximation.  The low-energy
excitations at moderate temperatures and small $U$ are quasiparticles
with a short lifetime.
As the temperature is lowered, coupling to evolving spin fluctuations
leads to the extinction of these quasiparticles, signaled by
a weak pseudogap in the density of states and by a
positive slope in Re $\Sigma ({\bf k}_F, \varepsilon)$ and a local
maximum in $|$Im $\Sigma ({\bf k}_F, \varepsilon)|$ at $\varepsilon = 0$.
We explain these results using a simple spin-fluctuation model.
\end{abstract}
\pacs{71.27.+a, 75.10.Lp, 75.30.Kz}
\narrowtext

The 2D single-band Hubbard model plays a central role in efforts to
understand the
behavior of electrons near the Fermi surface in the cuprate superconductors
and their parent compounds \cite{dagotto}.  The model is characterized by a
nearest-neighbor
hopping energy, $t$, and an on-site Coulomb energy, $U$.  For a half-filled
band, the $T = 0$ state is believed to be an antiferromagnetic insulator for
all $U > 0$ \cite{hirsch1,white1}.  For $T > 0$, the Mermin-Wagner theorem
precludes the existence of long-range AFM order in 2D, but
with strong coupling ($U \gtrsim 8t$) the low-temperature electronic state is
almost certainly a Mott insulator.   With $U = 4t$ and
$0.10t \lesssim T \lesssim 0.25t$,
conflicting results have been obtained from
quantum Monte Carlo (QMC) calculations of
the one-electron spectral weight function,
$A({\bf k}, \varepsilon)$,
depending on lattice size
and especially on the method used to extract $A({\bf k}, \varepsilon)$
from the Green's function $G({\bf k}, \tau)$ produced directly by QMC.
Spectral functions on the Fermi surface (FS)
produced by the maximum entropy technique
show a single peak whenever the lattice is
larger than the AFM correlation length, but
develop a pseudogap on smaller
lattices \cite{vwandw}.  In contrast, recent calculations
using the method of singular value decomposition yield a pseudogap
in both the spectral function on the FS and in the
total density of states,
$N(\varepsilon)  = N^{-1} \sum_{\bf k} A({\bf k}, \varepsilon)$,
even for lattices larger than the correlation length \cite{creffield}.

We report calculations of $A({\bf k}, \varepsilon)$ and
$N(\varepsilon)$ at half-filling and moderate $U$ ($ < 4.8t$) using the
fluctuation exchange approximation (FEA), a self-consistent conserving
approximation
that has been applied to the 2D Hubbard model in a number of recent papers
\cite{bickers1,tewordt,dwaves}.
In particular, the FEA has been used to argue for a d-wave superconducting
transition in the high-$T_c$ cuprates \cite{dwaves}; the need to evaluate
these claims adds to the importance of knowing
what the FEA predicts (rightly or wrongly) for the normal state at
half-filling.
Compared to QMC, the FEA has the disadvantage that it is inherently an
approximation, though imaginary-time Green's functions from the FEA
and QMC agree surprisingly well at half-filling and moderate $U$
\cite{bickers1}. For studying
the spectral function and DOS, the FEA has several important advantages
over QMC: (1) There is no inherently statistical error in the FEA results,
which removes most of the uncertainty in extracting $A({\bf k}, \varepsilon)$
from $G({\bf k}, \tau)$.  (2) FEA calculations are possible
for large enough lattices (typically $128 \times 128$) to ensure that the
profile
of $A({\bf k}, \varepsilon)$ represents the infinite lattice limit,
and does not in part reflect the small number of decay channels available to
a low-energy quasiparticle on a small lattice.  (3) The FEA can cover a wider
range of temperatures than are currently accessible to QMC.
(4) The FEA provides real-frequency self-energies and fluctuation propagators
that explain
the origin of structures in $A({\bf k}, \varepsilon)$ and
directly test the applicability of the
Fermi liquid picture.

We find that in the FEA a paramagnetic non-Fermi liquid state evolves with
  decreasing temperature.  The spectral functions on the FS
show no pseudogaps over the range of temperatures covered by
QMC calculations, but are exceptionally broad.  For momenta close to the
FS, spectral weight is shifted away from $\varepsilon = 0$,
producing a weak secondary maximum, which might suggest a ``shadow band''
\cite{dagotto},
although the real part of the denominator of the Green's function still has
only a single zero.  These features produce a weak pseudogap in the total DOS,
even though the FS spectral functions are single-peaked \cite{note1}.   For
momenta on
or near the FS, the real part of the
self-energy has an anomalous positive slope near $\varepsilon = 0$ and
the quasiparticle lifetime has a local minimum there.
A simple analytic model shows that these anomalies result from the
coupling of quasiparticles
to AFM spin fluctuations, without a phase transition to AFM order.
The process is anisotropic, beginning  at the $X$-point and spreading over
the FS.

The FEA for the self-energy in a paramagnetic state of the Hubbard model is
\cite{selfenergy}
\begin{eqnarray}
\Sigma({\bf r},\tau) & = & U^2 \left[ \chi_{ph}({\bf r},\tau) + T_{\rho\rho}
({\bf r},\tau) + T_{\sigma\sigma}({\bf r},\tau) \right] \,G({\bf r},\tau)
\nonumber \\
           & & \, \, \, \,  + U^2 T_{pp}({\bf r},\tau)G(-{\bf r},-\tau),
\label{fea}
\end{eqnarray}
where
$\chi_{ph}({\bf r},\tau) = - G({\bf r},\tau)\, G(-{\bf r},-\tau)$ and
$T_{\sigma\sigma}$, $T_{\rho\rho}$, and $T_{pp}$ are propagators for
spin fluctuations, density fluctauations, and singlet pair fluctuations.
For example, the spin-fluctuation propagator is
\begin{equation}
T_{\sigma\sigma}({\bf q},\omega_m)  =
 \  {{3}\over{2}} \ \ {{U \chi_{ph}({\bf q},\omega_m)^2 }\over
                             {1 - U \chi_{ph}({\bf q},\omega_m)}} .\\
\label{t_matrix}
\end{equation}
The Green's function is obtained from Dyson's equation,
\begin{equation}
G({\bf k},\varepsilon_n)^{-1} =G_o({\bf k},\varepsilon_n)^{-1} -
\Sigma({\bf k},\varepsilon_n),
\label{dyson}
\end{equation}
where $G_o({\bf k},\varepsilon_n)$ is the non-interacting Green's
function, and the self-consistent $\Sigma$ is found by
iteration.
 We have performed calculations on $128 \times 128$ lattices
  of {\bf k}-points \cite{why} with typically 512 Matsubara frequencies
using a   new massively parallel algorithm that
  treats high frequency contributions exactly \cite{deisz}.  This
greatly improved treatment of high-frequency information allows us
to obtain what we believe are the first meaningful results for the
FEA spectral functions of the Hubbard model at half-filling.
Numerical analytic continuation
of $\Sigma$ to the real-frequency axis is accomplished
using Pad\'e approximants \cite{vidberg}.

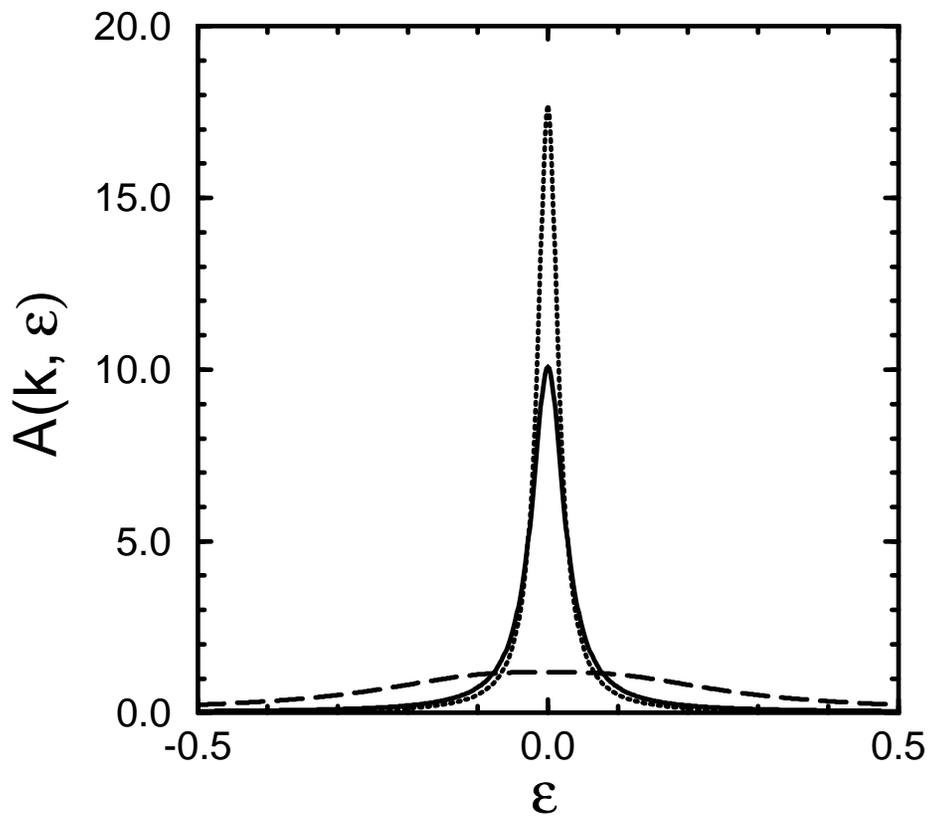
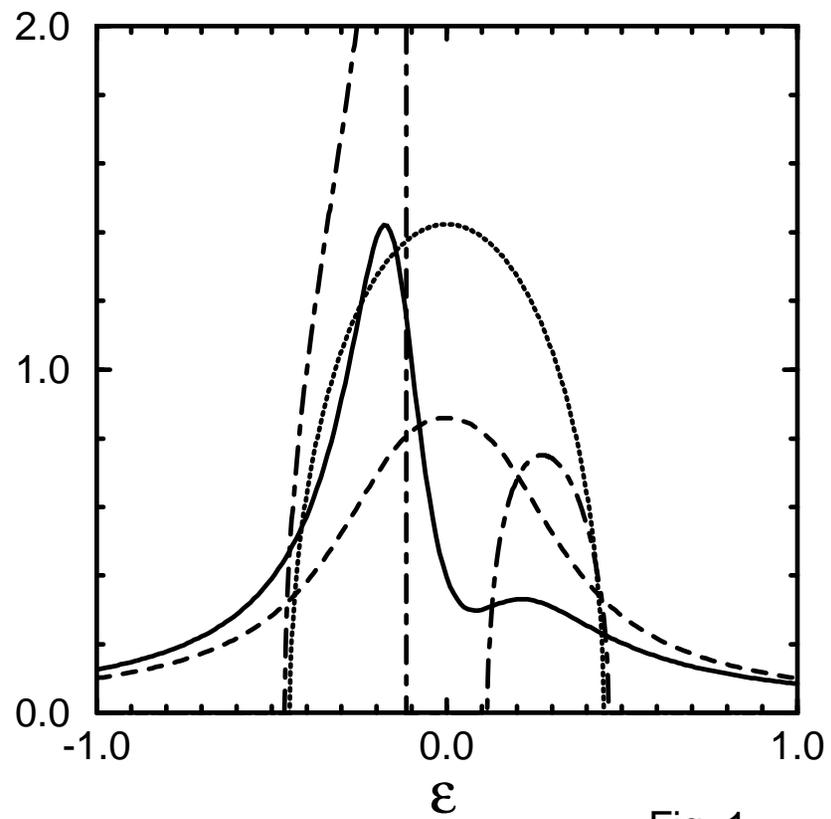
\begin{figure}
\caption{
\label{spec}  (left) Spectral functions calculated in the FEA
 at $T=0.10$, showing a dramatic reduction of spectral weight
 at the X-point (on the Fermi surface) with a modest increase in $U$
 between $T=0.10$ and $U=1.0$ (dotted) and $2.3$ (dashed).
 Also shown for $U=2.3$ (solid) is a calculation with
only the second order skeleton diagram.
 (right)
FEA spectral functions for ${\bf k}= (0.891 \pi, 0)$ (solid)
and ${\bf k}= (\pi, 0)$ (dashed) shown in comparison
 with those for the simple spin fluctuation model
 (long-dash-short-dash and dotted respectively) for $\tilde{t}_{sp}=0.05$.
}
\end{figure}

  The extinction of sharp quasiparticles at the $X$-point
  is evident from the spectral functions in Fig.~\ref{spec}.
  For $U=1.0$ and $T=0.1$, the FEA solution shows
  a sharp quasiparticle at $\varepsilon=0$,
 the Fermi energy (henceforth all energies are
  in units of $t$).
 While an increase in the interaction
  strength and thus an increase in the quasiparticle-quasiparticle
 scattering rate
  leads generally to a reduction in spectral weight at
  $\varepsilon=0$ and a broadening of the spectral function, the effect shown
  for $U=2.3$ in
  Fig. \ref{spec} is much larger than expected.

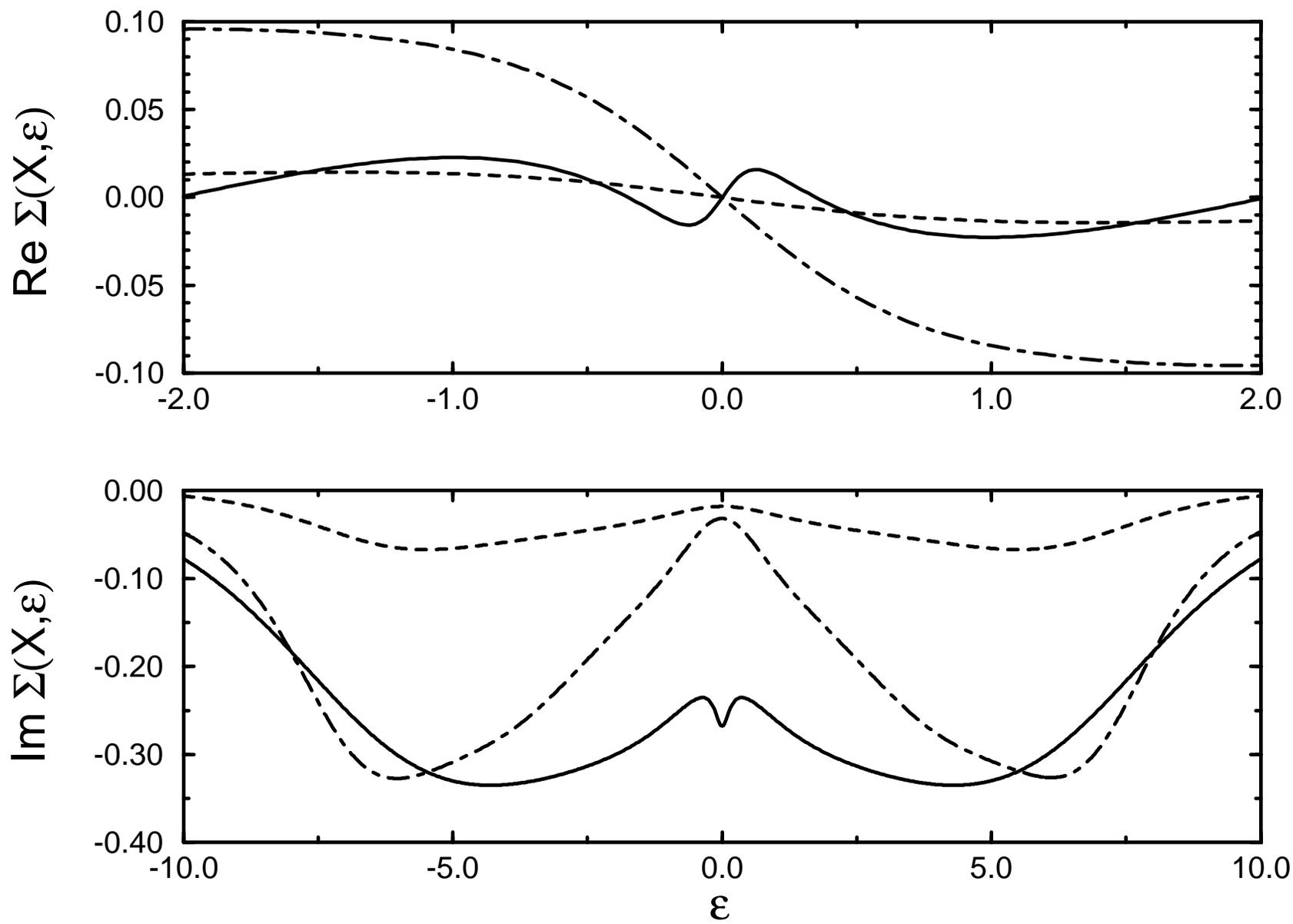
\begin{figure}
\caption{
\label{self-energy} The real and imaginary parts of the FEA self-energy
 at the X-point for $U=1.0$ (dashed) and $U=2.3$ (solid).
 Note the `inverted peak' at $\varepsilon=0$ in the imaginary part
 and the {\em positive} slope at low energy in the
 real part.  Also shown is
 the self-energy from a self-consistent calculation with only the
 second order skeleton diagram for $U=2.3$ (dash-dotted) which is
 qualitatively similar to the FEA for $U=1.0$.
}
\end{figure}

  To see this, focus on the
  self-energies at the $X$-point shown in Fig. \ref{self-energy}.
  Scaling of the $U=1$ result by $U^2$ leads to an Im $\Sigma$
  comparable to the $U=2.3$ result at high frequency,
  but smaller than the true self-energy for
 $\varepsilon \approx 0$ by a factor of $\approx 2.5$.
  The FEA self-energy at low energy
  is inconsistent with Fermi liquid theory:
  an anomalous maximum is clearly evident
  at  $\varepsilon = 0$ in $| {\rm Im} \Sigma |$,
   and Re~$\Sigma$ at $\varepsilon = 0$ has a positive
  slope which is inconsistent with the interpretation
  of $(1 - \partial \, {\rm Re} \Sigma({\bf k}_F, \varepsilon) \, /
   \, \partial \, \varepsilon \, |_{\varepsilon=0})^{-1}$ as the
  quasiparticle pole weight. For the same $U$ a
  self-consistent calculation with only the second-order skeleton
  diagram yields a sharp spectral function and self-energies
  without these anomalies.

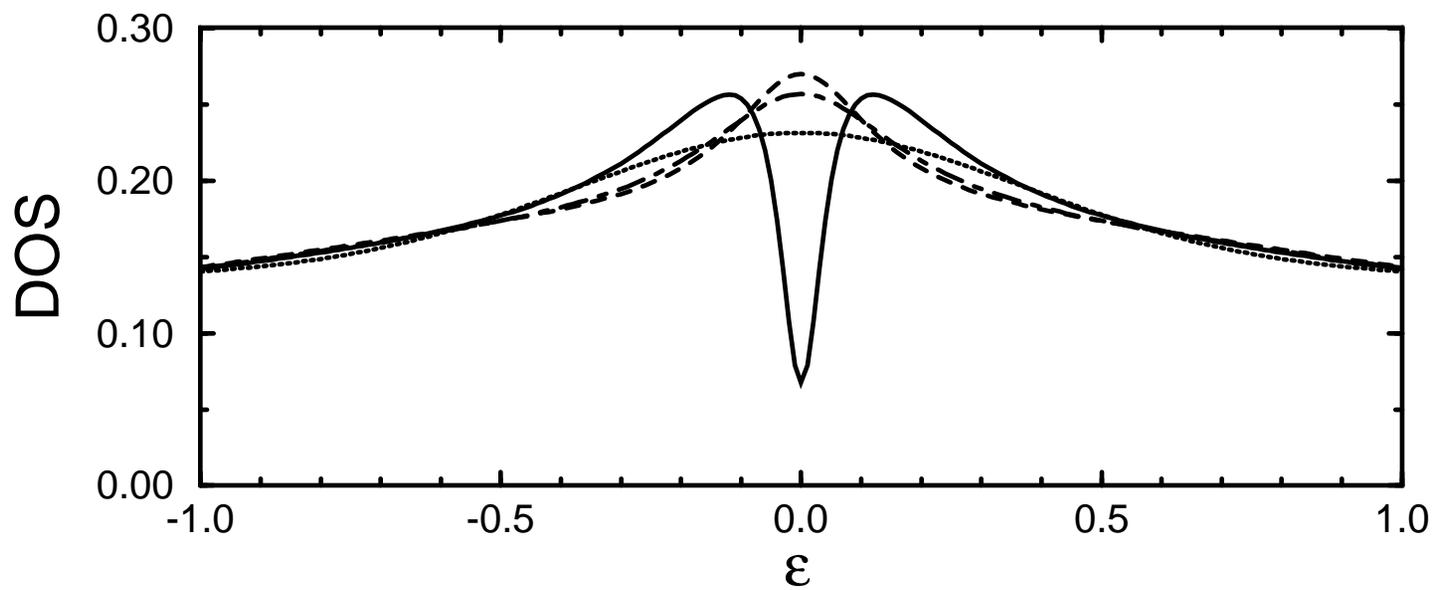
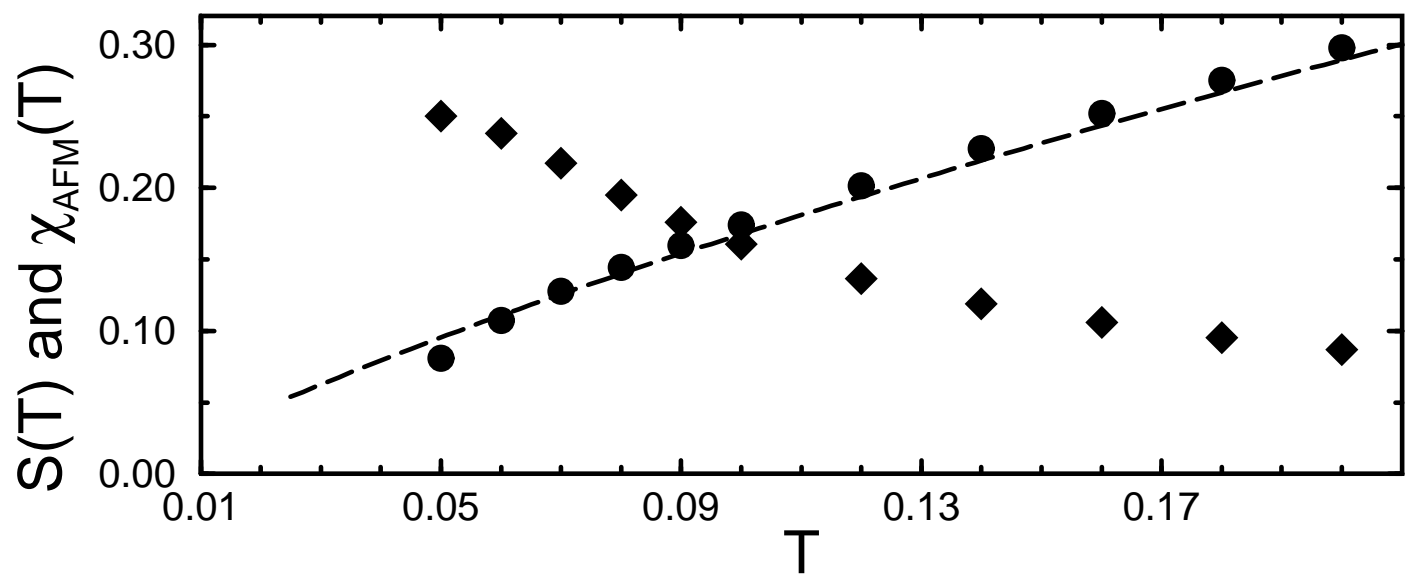
\begin{figure}
\caption{
\label{pseudo}
(top) The density of states as a function of energy for $U=1.57$ and
 $T= 0.16$ (dashed), $0.14$ (dash-dot), $0.07$ (dotted)
and $0.05$ (solid) showing the
 evolution of a weak pseudogap with decreasing temperature.
(bottom) The self-consistent AFM spin susceptibility (diamonds) and
entropy $(\bullet)$ as a function of temperature for $U=1.57$.
Also shown is the entropy for for $U=0$ (dashed). Note that for
clarity, $\chi_{AFM}(T)$ has been plotted as $\chi_{AFM}(T)/4
\chi_{AFM}(0.05)$.
There is no sign of an SDW
with ${\bf q} = (\pi, \pi)$ in the full spin-susceptibility.
}
\end{figure}

  The dramatic loss of spectral weight at the Fermi surface is reflected in
  the formation of a weak pseudogap in the total density of states as a
function
 of
  temperature as shown for $U=1.57$ in Fig. \ref{pseudo}.
  The anomalies in $\Sigma$ first appear at the
  van Hove critical points, and spread as $T$ decreases;
  for a range of $U$ and $T$, the anomalous excitations
  near the $X$-points coexist with
  more conventional quasiparticles
  on the rest of the Fermi surface.

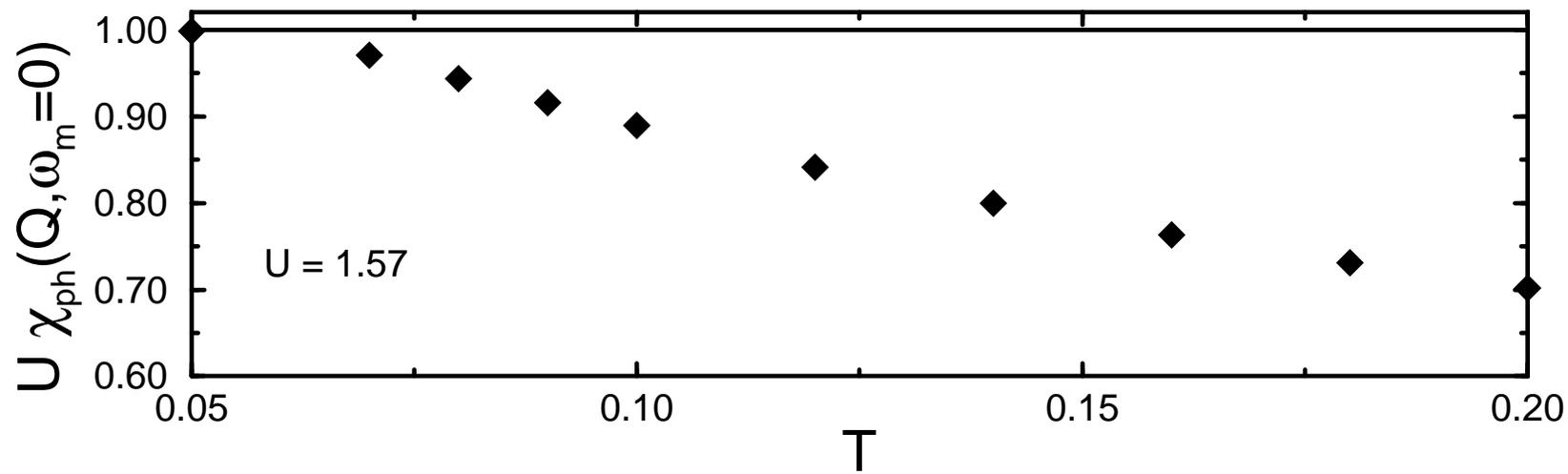
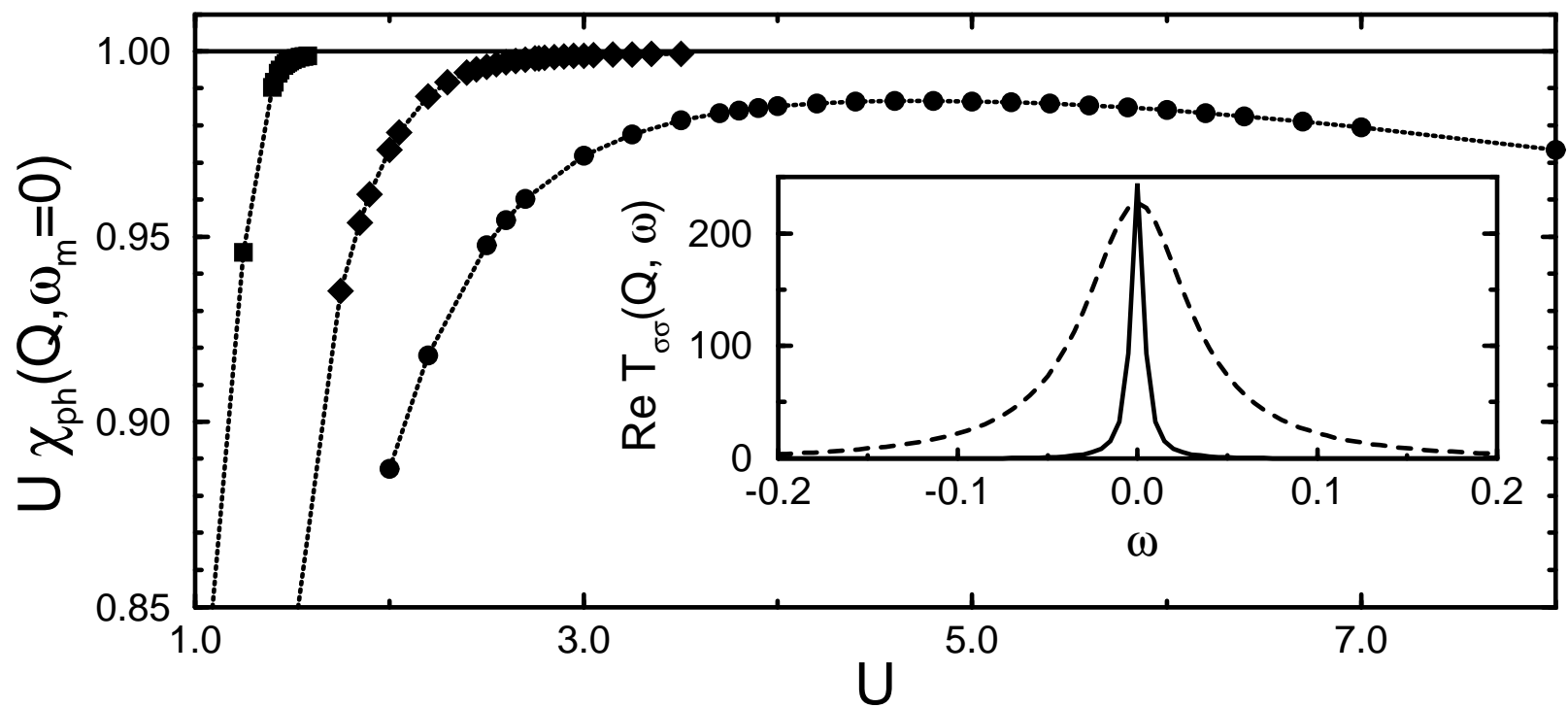
\begin{figure}
\caption{
\label{chi_max}  The self-consistent
$U \chi({\bf Q}, 0)$ (top) for $U=1.57$ as
a function of $T$ and (bottom) for $T=0.05 $ (squares),
$0.10$ (diamonds), and $0.15 \ (\bullet)$ as a function
of $U$. The inset shows analytic continuations of the
self-consistent $T$-matrix
Re~$T_{\sigma \sigma} ({\bf Q}, \omega)$ for $T=0.10$ and $U=2.7$ (solid)
and for $T=0.15$ and $U = 4.8$ ($\times 10$ and dashed).
The former shows a pseudogap in the density of states
but the latter does not.
}
\end{figure}

  The redistribution of spectral weight away from the Fermi energy
  is a consequence of strong antiferromagnetic spin correlations,
  signaled by the growth of  sharp structure  in the spin-fluctuation
$T$-matrix
  $T_{\sigma\sigma}$ when $U\chi_{ph}({\bf Q},0)$ approaches unity
  at ${\bf Q} = (\pi,\pi)$.
In a Hartree-Fock theory,  $U\chi_{ph}({\bf Q},0)$ reaches
  unity at a sufficiently low temperature,
  signaling a transition to AFM order.   Within the FEA,  the self-consistently
  determined quasiparticle lifetime regulates the growth of $U\chi_{ph}({\bf
Q},
0)$.
  As shown in Fig.~\ref{chi_max}, $U \; \chi_{ph}({\bf Q},0)$ closely
  approaches unity
  with decreasing temperature at fixed $U$, and with increasing
  $U$ for fixed $T$.   As the temperature is decreased
 for fixed $U$ a sharply peaked structure evolves in
 $T_{\sigma\sigma}({\bf q}, \omega_m)$
 within $|\Delta q| \approx 0.05 \pi$  of ${\bf Q}$
 (for $U = 2.7$ and $T=0.10$) and, to an excellent
 approximation, at $\omega_m = 0$.  For real frequencies,
 $T_{\sigma \sigma}( {\bf Q}, \omega)$ has sharply peaked
 structure as shown in Fig. \ref{chi_max}.
 At higher temperatures, as shown for $T=0.15$, temperature and correlations
 broaden the peak in  $U\chi_{ph}({\bf Q},0)$ and hence in
 Re~$T_{\sigma \sigma}$ (see Fig. \ref{chi_max} inset) sufficiently that
 a region of positive slope in
 Re $\Sigma({\bf k}_F, \varepsilon)$ at $\varepsilon=0$ is not observed for any
 $U$ and only a weak maximum at $U \approx 4.8$ is evident in
 $U\chi_{\rm ph}({\bf Q},0)$ as a function of $U$.

  At $T=0.1$, further increases
 in $U$ lead to neither a positive slope in Re $\Sigma$
 nor sharp structure in Im $\Sigma$ at $\varepsilon \approx 0$,
 perhaps because the large quasiparticle width `smears' the
 Fermi surface and reduces the effect of nesting
 and van Hove critical points.
 However, $U\chi_{ph}({\bf Q},0)$ continues to approach unity monotonically
 up to the largest $U$ $(\approx 3.6)$ for which we have solutions at $T=0.1$.
 Hence it is the sharpness of $U\chi_{ph}({\bf Q},0)$ (and $T_{\sigma \sigma}$)
 and not simply its size that leads to the observed anomalies.

The entropy $S$ and AFM susceptibility $\chi_{AFM}$ \cite{response} as a
function of temperature are shown in Fig. \ref{pseudo} for $U = 1.57$.
At high temperatures, $S(T)$ tracks the $U=0$ entropy.
As the pseudogap opens the entropy turns down,
signaling the loss of quasiparticle states.
For $T \approx 0.05$, $\chi_{AFM}$ shows only a modest enhancement
and the paramagnetic state remains stable, as required in 2D.
The nearly singular behavior of
$T_{\sigma\sigma}({\bf Q},0)$ is offset by the vertex corrections
included in a fully conserving description,
as suggested by previous calculations \cite{bickers2}.

 To understand better the self-consistent self-energy,
 we appeal to a simple model motivated by a calculation of the fluctuation
 conductivity above $T_c$ in superconductors \cite{patton}.
 The anomalies in $\Sigma$ are always accompanied by a
 $T_{\sigma\sigma}({\bf q},\omega_m)$ that
 is strongly peaked near
 ${\bf q} = {\bf Q}$ and $\omega_m =0$ \cite{schrieffer}.
 Assuming that the spin-fluctuation contribution to $\Sigma$ dominates,
\begin{equation}
\Sigma({\bf k},\varepsilon_n) \approx
G({\bf k} + {\bf Q},\epsilon_n)\, \tilde{t}_{sp},
\label{t_approximation}
\end{equation}
where $\tilde{t}_{sp}$ is proportional to
the weight of the peak in $T_{\sigma\sigma}$ within a reciprocal
correlation length $\xi^{-1}$ of ${\bf Q}$,
\begin{equation}
\tilde{t}_{sp} = {{U^2 T}\over{N}} \sum_{|{\bf q}| < 1/\xi}
T_{\sigma\sigma}({\bf Q} + {\bf q},\omega_m = 0).
\label{t}
\end{equation}
Eqs.\ (\ref{t_approximation}) and (\ref{t}), together with Dyson's equation,
can be solved directly.
For $\varepsilon \approx 0$ the slope of Re $\Sigma(\varepsilon)$ is 1/2.
In the full calculation, the slope
is also positive when $T_{\sigma \sigma}$ is sharply peaked, but
generally differs from $1/2$ due to other contributions to the
self-energy not included in Eq. \ref{t_approximation}.  We observe that
for slopes greater than unity, multiple quasiparticle peaks can
appear in the spectral function, corresponding to multiple solutions
of $ \varepsilon- \epsilon_{\bf k}
- \hbox{Re} \; \Sigma({\bf k},\varepsilon)  = 0$ for
a given wavevector.  For $T \geq 0.05$ we have not observed such a
splitting of the band, in contrast to at least two non-self-consistent
calculations describing the effects of strong antiferromagnetic
correlations on quasiparticle properties
\cite{kampf,deVries}.

The single-particle spectral function for the model
is nonzero only for
$ 0 < \varepsilon_- \varepsilon_+  < 4 \tilde{t}_{sp} $,
and given by
\begin{equation}
A({\bf k},\varepsilon)  =  {{1}\over{\pi \tilde{t}_{sp}}}
 \ \sqrt{{\varepsilon_+} \over{\varepsilon_-} } \
\sqrt{ \tilde{t}_{sp} - \varepsilon_- \varepsilon_+ / 4 },
\end{equation}
where $\varepsilon_- = \varepsilon - \epsilon_{{\bf k}}$ and
$\varepsilon_+
= \varepsilon - \epsilon_{{\bf k} + {\bf Q}}
= \varepsilon +\epsilon_{\bf k}$.
This spectral function shares two important features with the full
calculation.  First, spectral weight at $\epsilon_{\bf k}$ is shifted
away from $\varepsilon = 0$, with the non-interacting delta function
peak becoming a square root singularity truncated at $\epsilon_{\bf k}$.
This transfer of spectral weight from $\varepsilon=0$ leads to
the formation of a pseudogap in the total density of states similar
to the FEA pseudogap shown in Fig.~\ref{pseudo}.
Second, an additional peak forms at $\epsilon_{{\bf k} + {\bf Q}}$
as expected from dynamical coupling to antiferromagnetic
spin fluctuations. We emphasize that $\epsilon_{{\bf k} + {\bf Q}}
- \epsilon_{\bf k} - \hbox{Re}\;\Sigma({\bf k},\epsilon_{{\bf k} +
{\bf Q}}) \ne 0$ so that this second peak is not a second
quasiparticle solution.  Spectral functions for $\tilde{t}_{sp} = 0.05$,
a value consistent with $T_{\sigma \sigma}({\bf
q},\omega_m)$ for $U=2.7$ and $T=0.10$, are shown together
with those from the full FEA calculation in Fig. \ref{spec}. Given
the simplicity of the approximation, the agreement is good.
Results obtained from analytic continuation of quantum Monte Carlo
data at low temperature ($T=0.1$) and moderate coupling ($U=4$)
appear consistent with these general
features \cite{white2}.

In summary, the FEA yields thermally broadened
quasiparticles at high $T$ for any $U \lesssim 8t$.
With decreasing temperature and for a modest $U$,
a sharply-peaked spin fluctuation propagator develops, whose
coupling to the evolving quasiparticle excitations leads to
a radical breakdown of Fermi liquid theory and to a weak
pseudogap in the single-particle DOS, in the absence of
AFM order or of a gap in the spectral functions on the
Fermi surface.

This work was supported in part by a grant of computer time
from the Office of Naval Research and the DoD HPC
Shared Resource Centers: Naval Research Laboratory Connection Machine
facility CM-5; Army High Performance Computing Research Center,
under the auspices of Army Research Office contract number DAAL03-89-C-0038
with the University of Minnesota.

\begin{references}
\bibitem{why} A fine mesh is required to represent accurately sharp
      structure in the spin-fluctuation
      propagator  with decreasing temperature.
\end{references}
\end{document}